\documentclass[prl,preprint,aps,showpacs]{revtex4}

\usepackage{graphicx}
\usepackage{amssymb}
\usepackage{epstopdf}
\begin{document}
\renewcommand{\thefootnote}{\fnsymbol{footnote}}
\renewcommand{\theequation}{\arabic{section}.\arabic{equation}}

\title{A new water anomaly: the temperature dependence of the proton mean 
kinetic energy}

\author{D.~Flamini}
\author{M. A. ~Ricci}
\author{F.~Bruni$^{\star}$}
\affiliation{Dipartimento di Fisica "E.~Amaldi", Universit\`a degli Studi di Roma Tre,\
Via della Vasca Navale 84, 00146 Roma, Italy}
\footnotetext[1]{Author to whom correspondence should be addressed. E-mail: bruni@fis.uniroma3.it}

\begin{abstract}
The mean kinetic energy of protons in water is determined by Deep Inelastic 
Neutron Scattering experiments, performed above and below the temperature 
of maximum density and in the supercooled phase. The temperature dependence 
of this energy shows an anomalous behavior, as it occurs for many water 
properties. In particular two regions of maximum kinetic energy are 
identified: the first one, in the supercooled phase in the range 
269 K - 272 K, and a second one above 273 K. In both these regions the 
measured proton kinetic energy exceedes the theoretical prediction based on 
a semi-classical model. Noteworthy, the proton mean kinetic energy has a 
maximum at 277 K, the temperature of the maximum density of water. In the 
supercooled metastable phase  the measured mean kinetic energy and the proton 
momentum distribution clearly indicate proton delocalization between two 
H-bonded oxygens.
\end{abstract}
\date{\today}
 \maketitle

Among liquids, water is undoubtely the most studied, as it is the medium of life and 
chemistry on our planet, and its puzzling properties are a challenge for 
science \cite{Debene,PCCP}. Yet water quantum behavior has received less attention than 
its classical thermodynamic properties, in spite of the modest cooling needed to 
evidence quantum effects in water. These have been recently suggested to determine 
the water density maximum \cite{Deeney} and shown to influence the molecular 
geometry \cite{Soper} and in particular the OH bond-length. Quantum effects have 
also been evidenced in the short time dynamics of protons along the OH 
bond in supercooled \cite{noi} and confined water \cite{Garbuio,Bruni, Pagnotta}, 
by Deep Inelastic Neutron Scattering (DINS) \cite{advphys} experiments and computer 
simulations \cite{Morrone1,Morrone2}. Here we report DINS measurements on bulk water 
in the stable phase around the density maximum and in the supercooled metastable 
regime, showing clear evidence for quantum effects on the momentum distribution 
$n(p)$ and mean kinetic energy $\langle E_K \rangle$ of water protons. 

At present DINS is the only technique which determines the momentum distribution and 
mean kinetic energy of a single particle in condensed matter. It is based 
on neutron scattering measurements at high energy (1 eV $<\hbar\omega<$ 200 eV) and 
momentum (120\AA$^{-1}< \hbar q <$  300\AA$^{-1}$) transfers, thus probing both the 
short-time dynamics and local environment of the particle. The high energy and wave 
vector transfers achieved allow the scattering process to be described within the 
framework of the Impulse Approximations (IA) \cite{West}. This assumes that neutron 
scattering is incoherent and occurring within time scales much shorter than the 
typical relaxation times of the collective excitations of the system. In these conditions 
the struck particle recoils freely from the collision, with inter-particle interaction 
in the final state being negligible (i.e. the wave-function of the particle in its 
final state assumed to be a plane wave). In a molecular system, as for instance water, 
the contribution to the scattering cross section of protons can be easily distinguished 
from that of oxygens, due to the large mass difference. The scattering cross section 
is then expressed in terms of the single proton momentum distribution $n(p)$, whose variance 
is directely related to its mean kinetic energy $\langle E_k \rangle$. These 
quantities provide a 
richness of information about the potential surface that the proton experiences 
\cite{Reiter}, including the effects of hydrogen bonding, thus complementing microscopic 
structural studies \cite{Soper,Botti,Ricci} and allowing a direct comparison with 
quantum Monte Carlo simulations \cite{Morrone1,Morrone2}.

DINS experiments have been carried out on the VESUVIO spectrometer \cite{advphys}, at 
the ISIS spallation neutron source (UK). The  samples were contained in the same disk 
shaped aluminum can (5 cm diameter, 1 mm thickness) with inner Teflon coating used in 
previous work on supercooled water \cite{noi}. Experimental data have been recorded 
at 64 fixed-angle detectors in the standard Resonance Detector configuration \cite{acqua} 
employing the Foil Cycling technique \cite{Erik}. Data have been corrected for multiple 
scattering, aluminium and oxygen signal, by using the same routines as in previous 
work \cite{noi}.
Within the IA framework the dynamical structure factor, $S_{IA}(\vec{q},\omega )$, 
is related to the $n(p)$ through the relation:

\begin{equation}
S_{IA}(\vec{q},\omega )= \int  n(\vec{p})\; \delta\left(\omega - \frac{\hbar q^{2}}{2M} - 
\frac{\vec{p}\cdot \vec{q}}{M}\right)\, d \vec{p}\;
\end{equation}

\noindent where $\hbar \omega$ is the energy transfer, $\frac{\hbar^2 q^{2}}{2M}$ 
is the recoil energy of the struck atom of mass $M$, and $q$ is the vavevector transfer. 
The standard deviation of the $n(p)$ function, $\sigma$, is related to $\langle E_k \rangle$, 
through the relation $\sigma^2= \frac{2 M}{3h^2}\langle E_k \rangle$.
The dynamical structure factor is then expressed in terms of a Neutron 
Compton Profile (NCP) \cite{advphys}: $F(y)=\frac{\hbar q}{M}S_{IA}(\vec{q},\omega )$, 
where $y=\frac{M}{\hbar q}\left( \omega -\frac{\hbar q^{2}}{2M}\right) $ is the West 
scaling variable \cite{advphys}. The NCP lineshape is convolved with the instrumental 
resolution function, and represented as a series expansion in Hermite polynomials:

\begin{equation}
F(y)=\frac{e^{-\frac{y^{2}}{2\sigma^{2}}}}{\sqrt{2\pi}\sigma}\left[1+\sum_{n}
\frac{c_{n}}{2^{2n}n!}H_{2n}\left(\frac{y}{\sqrt{2}\sigma}\right)\right]\label{GR}\;
\end{equation}

\noindent The coefficients $c_{n}$ and $\sigma$, appearing in the series expansion, 
have been determined by a least squares fitting procedure (see caption of Fig. 2), 
and small corrections due 
to deviations from the IA were taken into account \cite{advphys}. 
We did not attempt to correct for the so-called \textit{intensity deficit} seen for 
hydrogen relative to heavier nuclei \cite {cha97,sen05,rei05}. Altough the issue of the 
intensity deficit in DINS experiment is still debated, it should be noted that 
in systems where the $n(p)$ and the NCP are purely gaussian, such as ZrH$_{2}$, no 
distortion of the measured NCP is indeed observed \cite{eva96}. 
Finally the $n(p)$ is expressed by the expansion \cite{advphys}

\begin{equation}
n(p)= \frac{e^{-\frac{p^{2}}{2\sigma^{2}}}}{\left(\sqrt{2\pi}\sigma\right)^3} 
\cdot \sum_{n}c_n (-1)^n L^{\frac{1}{2}}_{n}\left(\frac{p^{2}}{2\sigma^{2}}\right)
\end{equation}

\noindent where $L^{\frac{1}{2}}_{n}$ are generalized Laguerre polynomials. Full 
details about data analysis may be found in ref. \onlinecite{advphys}.

Figure 1 shows the temperature dependence of proton's $\langle E_k \rangle$ measured 
in both stable and metastable phases of bulk water in the temperature range from 300 K 
to 269 K, along with the semi-classical prediction, $E_{sc}$. The temperature 
dependence of  $E_{sc}$ in the stable water phase, shown in Fig. 1 as a dotted line, 
has been derived in Ref. \onlinecite{Degiorgi}, taking into account translational, 
rotational, and vibrational contributions, including the optical spectroscopic data 
available in the literature. 
While data for ice \cite{ghiaccio} and water \cite{acqua} above 293 K, 
including supercritical states \cite{Degiorgi,Pantalei}, are satisfactorily 
described by the semi-classical prediction, data around the temperature of maximum 
density and in the supercooled phase show an excess of proton mean kinetic 
energy. In particular $\langle E_K \rangle$ shows 
two maxima: one at 277 K and the other in the supercooled phase in the range 
269 K - 271 K, with an excess of $\langle E_K \rangle$ with respect to $E_{sc}$  
of about 3 kJ/mol (30 meV) and 12 kJ/mol (120 meV), respectively.
The peculiar temperature dependence of $\langle E_k \rangle$, suggests that two distinct 
mechanisms may be considered below and above the melting temperature. As a matter 
of fact the excess of $\langle E_k \rangle$ in the stable water phase is moderate and its 
temperature dependence follows that of density \cite{NIST}, showing a 
maximum at the same temperature (see insert in Fig. 1). A correlation between 
density and $\langle E_k \rangle$, as measured by DINS, is not a novelty in principle, as it 
has already been observed in helium \cite{Bafile,Filabozzi} and explained in that case 
by using a harmonic model for the fluid. Nevertheless the peculiar density behavior 
of water and the temperature evolution of the H-bond network does not allow a similar 
model to work in this case. We suggest instead that the anomaly of $\langle E_k \rangle$ vs $T$ 
above 273 K may be explained as a further evidence for water structural anomalies, which 
manifest through the existence of a maximum of density and transport properties in 
the stable water phase \cite{debeneNature}. Within this hypothesis the maximum 
of $\langle E_k \rangle$ at 277.15 K, shown in Fig. 1, may be an indirect manifestation 
of the competition between zero point energy, $E_0$, and thermal fluctuations, which 
has been proposed as the quantum origin of the density maximum \cite{Deeney}. We notice 
however that quantum effects are not necessarily required to explain the existence of a 
maximum of density in water \cite{Voth}.

On the other side the huge increase of $\langle E_k \rangle$ in the metastable states of water, 
already observed in a recent publication \cite{noi} and confirmed by present new 
measurements at 272.15 and 272.95 K, seems to be directely related to the likely 
delocalization of protons along the H-bond. Fig. 2 shows the radial momentum distribution, 
 $4\pi p^2 n(p)$, at 277.15 and 271.15 K. This function at the lowest temperature 
shows a narrowing at low-$p$ and a shoulder at high-$p$ ($\sim$17\AA$^{-1}$), compared 
to that measured at 277.15 K, compatible with the transition from a single to a 
double well potential. The presence of a shoulder at high-$p$ in the radial momentum 
distribution has indeed been ascribed to coherent delocalization of protons over two 
sites of a double well potential \cite{Reiter} felt by the proton along the H-bond direction  
between two water molecules. In other words, a statistical ensamble of water molecules 
in which protons are localized in the vicinity of the covalently bonded oxygens is 
predicted to show a single maximum of $4\pi p^2 n(p)$ at $p\sim$ 6 \AA$^{-1}$, 
corresponding to the intramolecular O-H distance $d=\frac{2\pi}{p}\sim$1 \AA, as 
shown in Fig. 2 and confirmed by quantum mechanical simulations \cite{Morrone1,Morrone2}. 
Conversely, the appearence of a shoulder at $p\sim$ 17 \AA$^{-1}$ in the metastable 
supercooled phase indicates a delocalization of the proton over a distance 
$\Delta d= \frac{2\pi}{\Delta p} \sim 0.6$ \AA~ from the equilibrium position. 
We notice that $\langle E_k \rangle$ in the supercooled phase is comparable with the H-bond 
energy ($\sim$ 20 kJ/mol) and $\Delta d$ is ompatible with the width of the fluctuations 
of the H-bond length (width of the first intermolecular peak of the oxygen-hydrogen radial 
distribution function) and the oxygen-oxygen distance in supercooled 
water \cite{Botti}. Based on this observation, we proposed \cite{noi} the quantum 
excess of proton mean kinetic energy be correlated to the average distance between 
two first neighboring H-bonded oxygens in water. Such assignement was corroborated 
by similar evidences in confined water \cite{Garbuio,Ricci,Pagnotta}. Further 
evidences for a correlation between proton delocalization and distance between its 
neighboring oxygens are given in Ref. \onlinecite{Scheiner} and \onlinecite{Marx}, 
showing the changes of the potential landscape, energy barrier and wave-function of 
a proton as a function of this O-O distance.    

The present results show a clear and unexpected anomalous temperature 
dependence of the proton mean kinetic energy. Below 273 K, in the supercooled phase, 
this anomalous behavior is associated with a coherent delocalization of the 
proton between first neighbor oxygens. Above 273 K the temperature dependence 
of the proton mean kinetic energy resembles that of water density. 

\begin{acknowledgments}
The authors acknowledge useful discussion with C. Andreani, A. Pietropaolo, R. Senesi, 
and J. Mayers. We are grateful to M. Adams for helpful assistance and discussions 
during the experiment.  
This work has been performed within the Agreement No.01/9001 between STFC and 
CNR, concerning collaboration in scientific research at the spallation neutron 
source ISIS and with partial financial support of CNR. 
\end{acknowledgments}

\newpage
\section*{Figure Captions}
\begin{itemize}
\item{Figure 1:} Water proton mean kinetic energy $\langle E_K \rangle$ as a function of 
temperature. Open symbols have been used for water in the stable phases, solid symbols 
for the metastable supercooled phase, namely: ice \cite{ghiaccio} at 269 K (square); 
liquid stable water (circles). Water data above 293 K are taken from Ref. 
\onlinecite{acqua}, those below 272.15 K are from Ref. \onlinecite{noi}. Present 
data are reported for temperatures in the range 272.15 - 285.15 K. The dotted line 
corresponds to the semi-classical prediction \cite{Degiorgi}. The error bars are 
derived from the least squares fitting procedure used to derive $\langle E_K \rangle$ 
from the differential cross sections (see references \onlinecite{Garbuio,advphys}). 
The solid and dashed lines are guides for the eye.
In the insert $\langle E_K \rangle$ data in the stable liquid phase (circles and 
right axis) are reported in comparison with the density of water (line plus solid 
triangles and left axis) as a function of temperature.
\item{Figure 2:} Spherically averaged momentum distribution of water protons at 
the point of maximum density (present experiment, $T=$ 277.15 K), (dotted line) and 
in the metastable 
supercooled phase at $T=$ 271 K (solid line, from Ref. \onlinecite{noi}). Experimental 
uncertainties are less than $\pm $1 $\%$. Values for the fitting parameters (see Eq. 2-3) 
are as follows: $c_1$ is set to zero by definition,  the non-Gaussian coefficient, $c_2$, and 
$\sigma$ are (0.443 $\pm$ 0.008) and (6.05 $\pm$ 0.03)\AA$^{-1}$ at $T=$ 271 K; 
(0.148 $\pm$ 0.010) and (5.13 $\pm$ 0.02)\AA$^{-1}$ at $T=$ 277.15 K;  the coefficients 
$c_{n\geq 3}$ are found to be negligible.
\end{itemize}
 
 \newpage
\begin{figure}
\includegraphics [width= 12cm] {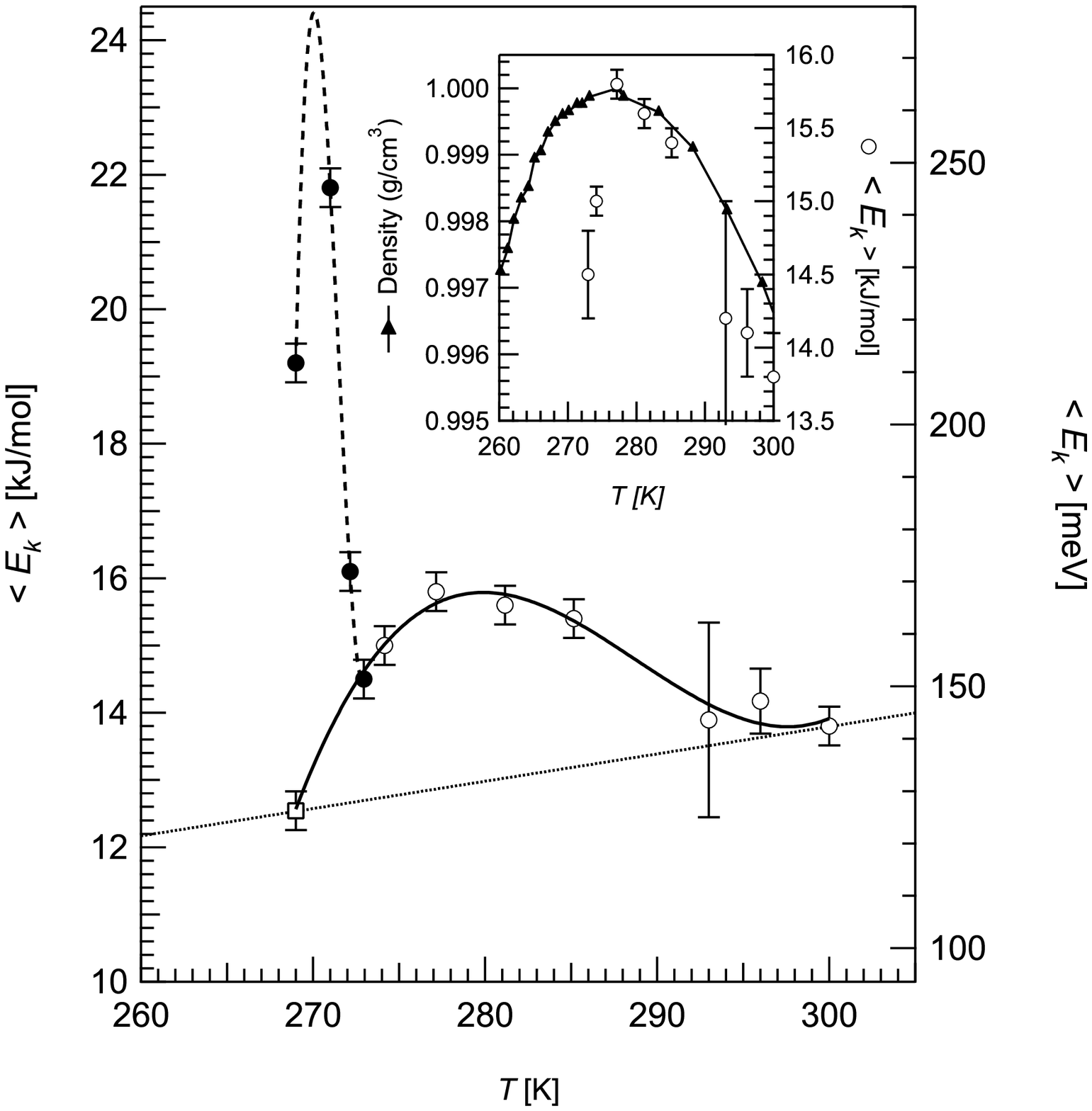}
\caption{}
\end{figure}

\begin{figure}
\includegraphics [width= 12cm] {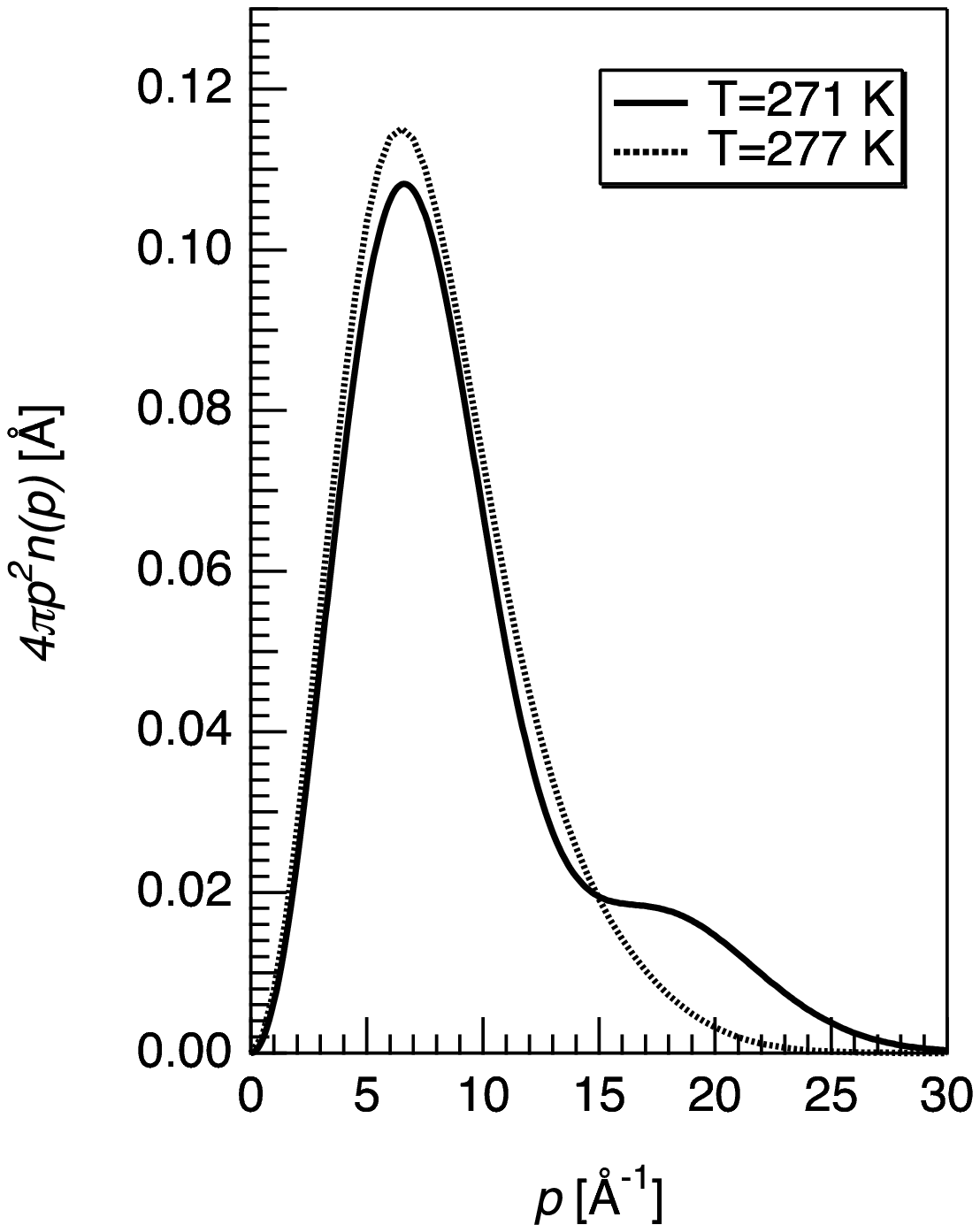}
\caption{}
\end{figure}

\end{document}